\documentclass[superscriptaddress,groupedaddress,nofootnoteinbib,12pt]{article}  

\usepackage{amsmath}
\usepackage{amsfonts}
\usepackage{authblk}
\usepackage{bbm}
\usepackage{color}
\usepackage[dvipsnames]{xcolor}
\usepackage{graphicx,graphics}
\usepackage{epstopdf}
\usepackage{caption}
\usepackage{subcaption}
\usepackage{float} 
\usepackage{pifont}
\usepackage{titlesec}
\usepackage{etoolbox}
\usepackage{indentfirst}
\usepackage{jcappub}
\usepackage{braket}

\def\bb{\begin{equation}}
\def\ee{\end{equation}}
\def\ba{\begin{array}}
\def\ea{\end{array}}
\def\babc{\begin{subequations}}
\def\eabc{\end{subequations}}

\def\5{\hspace*{5mm}}
\def\2{{\scriptstyle\frac12}}

\begin{document}

\begin{flushright}
NYU-TH-14/03/2020
\end{flushright}

\vspace{0.5in}

\begin{center}

\Large{\bf Conformal/Poincar\'e Coset,  Cosmology,  and
Descendants of Lovelock Terms}

\vspace{0.5in}

\large{Gregory Gabadadze and Giorgi Tukhashvili}

\vspace{0.2in}

\large{\it Center for Cosmology and Particle Physics, Department of Physics, \\
New York University, 726 Broadway,  New York, NY, 10003}

\vspace{0.3in}

\end{center}

We calculate six invariant terms of a gravitational field theory
that nonlinearly realizes the Conformal/Poincar\'e quotient, and reduce to the
known conformal Galileons in the limit  when only the conformal mode is kept.
Five of the six terms are regular coset terms, while  the sixth is a Wess-Zumino (WZ)
term that gives the well-known gravitational action for the trace anomaly.
The obtained terms can be embedded in a quantum
effective field theory (EFT)  without spoiling  their key features, although
at a cost of certain fine tunings. The additional massive modes that appear in the EFT
would have been  troublesome, however, for  sub-Planckian
curvatures their masses are (super)-Planckian and therefore the respective states
are outside of the EFT regime.  We discuss certain novel cosmological solution of
this theory and their validity within the EFT.  Furthermore, we show that the obtained
4D terms, except the WZ term,  can also be derived from higher dimensional
Lovelock terms by  reducing the latter to the genuinely four dimensional terms
according to a  well-defined algorithm.

\newpage



\section{\Large Introduction and summary}


Theories that modify gravity in the infrared, such as brane-induced gravity \cite{Dvali:2000hr} or
massive gravity \cite{deRham:2010ik,deRham:2010kj}, also exhibit interesting ultraviolet
properties as quantum effective field theories:  the terms that amend the Einstein-Hilbert (EH) action
in these theories --  e.g., the covariant graviton mass  term -- turn out  to contain  special
higher dimensional (irrelevant) operators at a scale much below the Planck
mass; while the emergence of this strong interaction scale is not unlike in massive
non-Abelian gauge theories \cite {Vainshtein:1971ip}, both
the dynamics  and open questions due to these higher dimensional  operators in massive
gravity are considerably richer, see, e.g.,
\cite {Vainshtein:1972sx,Deffayet:2001uk,ArkaniHamed:2002sp,Luty:2003vm,Adams:2006sv,Babichev:2013usa,Cheung:2016yqr,deRham:2017xox,deRham:2018qqo,Bonifacio:2019mgk,Ogawa:2019gjc}.

These  irrelevant operators describe nonlinear interactions of helicity-1 and helicity-0 components
of a massive spin-2 graviton. In particular, the helicity-0 interactions are described by the so-celled
Galileon field theories \cite {Luty:2003vm,Nicolis:2008in}.

In the absence of gravity the Galileon is an interesting effective field theory by its own.
If one were to start with a Galileon as a stand-alone scalar
field theory without gravity, then we know  that massive gravity  would provide
gravitational dressing of the Galileon.  In the dressed theory the Galileon
appears as a gauge mode of a tensor field.
In more physical terms, the Galileon particle is a Nambu-Goldstone (NG) mode that is
being  absorbed as a longitudinal mode  by a massive spin-2 state.\footnote{Note that this is
different from the direct covariantization  of the Galileon  \cite{Deffayet:2011gz}, where
the Galileon is an independent physical scalar degree of freedom coupled to gravity.}
Then, nonlinear interactions of the longitudinal model  are restricted   severely
by  the requirement   for the theory to propagate only  five physical
degrees of freedom of a massive spin-2 on an arbitrary background, and this requirement selects the Galileons
\cite{deRham:2010ik,deRham:2010kj,Mirbabayi:2011aa}.

There is another interesting class of derivatively interacting scalar field
theories, the conformal Galileons \cite{Nicolis:2008in}. They have been used in
cosmology to describe  alternatives for the early universe
\cite {Creminelli:2010ba,Creminelli:2012my,Pirtskhalava:2014esa}, where the
gravity was  introduced through direct covariantization of the
conformal Galileon  action, with the total action containing the physical
conformal Galileon field  alongside with a massless graviton.

It seems natural to ask  \cite{Gabadadze:2018hos}  whether  there  exists a gravitational
field theory  where  the conformal Gallileon  would be a NG mode (gauge mode),
like the ordinary Galileon is in massive gravity.

This question was addressed in  \cite{Gabadadze:2018hos} in 3D. The approach used
the observation that the conformal Galilleon terms emerge as coset  and Wess-Zumino (WZ) terms
in  a  non-gravitational theory,  with the conformal group spontaneously broken
to the Poincar\'e subgroup \cite {Goon:2012dy}. To obtain the gravitationally dressed
conformal Galileons  we considered a full gravitational Conformal/Poincar\'e coset in 3D.
This led to the well know action of  New Massive  Gravity  (NMG)
\cite{Bergshoeff:2009hq} and its generalization \cite{Sinha:2010ai},
both obtained in a local Weyl invariant form. The conformal Galileon itself
emerged as a NG mode of the broken scale invariance and
as a gauge mode of the local Weyl symmetry.

The next natural question appears to be this: what is a 4D gravitational theory that contains
conformal Galileons as NG/gauge modes? This seems to be a worthy question to address
since massive gravity cannot give an answer to it,  and a  4D generalization
of NMG is not known. Besides, the sought theory might  be expected to inherit some interesting
cosmological solutions of the conformal Galileon.

In the present work we will build such a 4D theory using the coset construction.  The coset will give
us six special terms: the first one is the Weyl  invariant version of the cosmological constant,
the second is the Weyl invariant EH term, the third one is the Gauss-Bonnet (GB) term,
the fourth  is a special combination of  invariant cubic curvature  terms,
the fifth is  a special combination of invariant quartic curvature  terms, and,
the sixth one emerges as a WZ term describing the action for the
conformal anomaly  \cite{Riegert:1984kt} (see also
 \cite{Komargodski:2011vj}). All  these terms, except the fourth and fifth, are well known.
Their defining feature  is that their conformal mode
exactly reproduces all the known conformal  Galileons. Hence, the equations of motion
of the theory -- once reduced to the conformal mode -- are necessarily of the second
order. Since the conformal mode is a scale factor in homogeneous and isotropic cosmology,
this theory gives novel, potentially viable cosmological solutions, as we will show.

The obtained action should be viewed  as part of  a certain quantum effective field theory (EFT)
with an infinite number of other terms. The EFT  however, can
only contain additional terms proportional to at least one power
of the Weyl tensor to preserve the above described properties. This requires
fine tunings of the counter-terms in the standard renormalization procedure.
The obtained theory would harbor additional degrees  of freedom due to the higher
derivative terms, but  their masses are at or above the cutoff of EFT
for reasonable values of other parameters. Thus, the theory albeit being fine tuned,
is otherwise a good EFT. \footnote{Fine tuning without symmetry is
viewed as a deficiency by the authors.}

Last but not least,  we show that five out of  the six new terms  can be
obtained from the higher dimensional Lovelock terms via  certain dimensional reduction
across various dimensions.  Only the sixth,  the  WZ term,  can not  be obtained  through
this procedure,  since it is not Weyl invariant.

Interestingly,  D. Glavan and Chunshan Lin \cite {Glavan:2019inb}  proposed a certain
continuation of the 5D GB term to four-dimensions. The obtained terms have their merits,
however, are not truly four-dimensional \cite {Gurses:2020ofy,Bonifacio:2020vbk}.
Our procedure bears a formal resemblance to that  of \cite {Glavan:2019inb},
but is  both conceptually and technically different:
the reduction of the Lovelock terms we perform
gives  local well defined terms in lower dimensions; for instance,
the 5D GB term reduces to a 3D local term of NMG,  while the 6D Lovelock term to
a pure 4D term belonging to the coset, as shown in section 5.

Throughout the  paper  we use the following conventions and notations: $\delta^{\mu_1 \cdots \mu_n}_{\nu_1 \cdots \nu_n} = \delta^{\mu_1}_{\nu_1} \cdots \delta^{\mu_n}_{\nu_n} \pm \text{permutations}$. $D$ stands for both, the generator of dilatations and covariant derivative. The Planck mass will be set to one, unless it is shown explicitly. The sign, $\simeq$, means ``equals up to a total derivative".  The Riemann tensor is ${R^\rho}_{\mu \sigma \nu} = \partial_\sigma \Gamma^\rho_{\mu \nu} + \cdots$. $R^{n+1}_{\mu \nu} \equiv R_\mu^{\rho_1} R_{\rho_1}^{\rho_2} \cdots R_{\rho_n \nu }$ and $\left[ R^n \right] \equiv R_{\rho_2}^{\rho_1} R_{\rho_3}^{\rho_2} \cdots R^{\rho_n}_{\rho_1}$. The total derivatives for the helicity-0 mode  start with the terms normalized as
$L_n^{TD} = \left( \partial^2 \pi \right)^n + \cdots$. The Schouten tensor is normalized as follows:
$S^\mu_\nu = \frac{1}{n-2} ( R^\mu_\nu - \frac{R}{2 ( n-1)} \delta^\mu_\nu )\,.$

\vspace{0.2in}


\section{\Large $SO(n,2)$ broken to  $ISO(n-1,1)$}\label{sec_coset}


\vspace{0.1in}

Let us briefly summarize the coset formalism \cite {Callan:1969sn} adopted for  the conformal group
\cite{Ivanov:1975zq,Borisov:1974bn}. One starts  in $n$ space-time dimensions and postulates the conformal group
to be spontaneously broken  down to its Poincar\'e subgroup. Conformal algebra in $n$ dimensions is realized by the $n(n+1)/2$ Poincar\'e generators $(P_a,~J_{ab})$, plus  $n$ generators of special conformal transformations, $K_a$, plus one generator of dilatations, $D$. These generators satisfy the following standard commutation relations:
\begin{align}
[P_a, D] = P_a, ~~~~~~~~~~~~~~~~~~~~~~~~  & [D,K_a] = K_a, \\
[J_{ab}, K_c] = \eta_{ac} K_b - \eta_{bc} K_a, ~~~~~~~ & [K_a , P_b ] = 2 J_{ab} - 2 \eta_{ab} D, \\
[J_{ab}, P_c] = \eta_{ac} P_b - \eta_{bc} P_a,  ~~~~~~~~~& [J_{ab},J_{cd}] = \eta_{ac} J_{bd} - \eta_{bc} J_{ad} + \eta_{bd} J_{ac} - \eta_{ad} J_{bc}\,.
\end{align}
Since $[K_a , P_b ] \propto - 2 \eta_{ab} D$ not all the NG fields of the coset  are independent; the Inverse Higgs Constraint (IHC) \cite{Ivanov:1975zq} will enable us to eliminate the
NG's related to broken $K_a$'s in favor of the NG  related to broken $D$.


\subsection{\large Flat space-time}

As a warmup we start with a  flat space-time metric, $\eta_{\mu\nu}$.  A convenient
parametrization for the $SO(n,2)/ISO(n -1,1)$ coset element is given by
\bb
\Sigma = e^{\pi D} e^{\xi^a K_a}\,.
\ee
One can construct the Maurier-Cartan one-form as follows:
\bb\label{flat_coset}
\Sigma^{-1} \left( d + \delta^a P_a \right) \Sigma = E^a P_a + \omega_K^a K_a + \omega_D D - \frac12 \omega_J^{ab} J_{ab}\,.
\ee
Here $\delta ^a \equiv \delta^a_\mu dx^\mu$, and the extra piece in the parenthesis
on the  left hand side (LHS)   is introduced for further convenience (it can be traded for an extra factor of
$e^{x^a P_a}$ in the coset element
in an equivalent approach). The expressions for the one forms in front of the
generators  on the RHS can be calculated using conformal algebra given above:
\bb\label{fl_cos_el_1}
E^a = e^\pi \delta^a\,,
\ee
\bb\label{fl_cos_el_2}
\omega_D = d \pi + 2 E^a \xi_a\,,
\ee
\bb\label{fl_cos_el_3}
\omega_K^a = d \xi^a - \xi^2 E^a + \xi^a \omega_D\,,
\ee
\bb
\omega_J^{ab} = -2 E^a \xi^b + 2 E^b \xi^a\,.
\ee
Here $\xi^a$ is a zero form Lorentz vector. We make a distinction between the Latin and Greek indices
in preparation to introduce  a non-trivial metric for a general pseudo-Riemannian space-time manifold;
the Latin indices will be used for tangent  space-times.


\subsection{\large Dynamical metric}


To covariantize the coset in the first order formalism we gauge the translations and rotations \cite{Delacretaz:2014oxa}. To gauge translations we simply replace $\delta^a$ by the field $e^a$ on the LHS of (\ref{flat_coset}), but for  gauging
rotations an extra piece is needed:
\bb\label{covariant_coset}
\Sigma^{-1} \left( d + e^a P_a - \frac12 \omega^{ab} J_{ab} \right) \Sigma = E^a P_a + \Omega_K^a K_a + \Omega_D D - \frac12 \Omega_J^{ab} J_{ab}\,.
\ee
A straightforward calculation leads to:
\bb
E^a = e^\pi e^a,
\ee
\bb
\Omega_D = D\pi + 2 E^a \xi_a,
\ee
\bb
\Omega_K^a = D \xi^a - \xi^2 E^a + \xi^a \Omega_D,
\ee
\bb
\Omega_J^{ab} = \omega^{ab} -2 E^a \xi^b + 2 E^b \xi^a \,.
\ee
Here the gauge field $e^a$ can be interpreted as an n-bein and $\omega^{ab}$ as a
spin connection of pure spin-2 field, $D$ corresponds to covariant derivative with respect to $\omega^{ab}$, i.e. $D = d + \omega$. Since we are working with  a dynamical pseudo-Riemannian manifold it is natural to construct
the curvature two form:
\bb
\mathcal{R}^{ab} = d \Omega_J^{ab} + \Omega_J^{ac} \wedge {{\Omega_J}_c}^b = R^{ab} + 2 E^a \wedge \Omega_K^b + 2 \Omega_K^a \wedge E^b\,,
\ee
\bb
R^{ab} = d \omega^{ab} + \omega^{ac} \wedge {{\omega}_c}^b\,.
\ee
 In the next section  we will use some of  these elements, but not all,  to build an effective action.



\section{\Large 4D Conformal Galileons and their Embedding}


The goal of this section is to show that the 4D conformal Galileon
describes a theory of the conformal mode of a diffeomorphism
invariant gravitational theory, made out of the Conformal/Poincar\'e coset
and WZ  terms.

\subsection{\large Flat Background Metric}\label{sec_review}


Ref.  \cite{Goon:2012dy}  showed that  the flat space
conformal Galileons can be constructed using the flat space Conformal/Poincare coset.
For convenience we will briefly summarize this remarkable result below  before introducing
the gravitational field in the next subsection.

Among the three one-forms in
(\ref{fl_cos_el_1})-(\ref{fl_cos_el_3}) only two carry a single local Lorentz
index.  Therefore only those terms, (\ref{fl_cos_el_1}) and (\ref{fl_cos_el_3}),
can be used to construct the four-form actions without invoking additional
covariant derivatives:
\bb
A_0 = \int_{\mathcal{M}_4} \varepsilon_{abcd} \,E^a \wedge E^b \wedge E^c \wedge E^d,
\ee
\bb
A_1 = \int_{\mathcal{M}_4} \varepsilon_{abcd} \,E^a \wedge E^b \wedge E^c \wedge \omega_K^d,
\ee
\bb
A_3 = \int_{\mathcal{M}_4} \varepsilon_{abcd} \,E^a \wedge \omega_K^b \wedge \omega_K^c \wedge \omega_K^d,
\ee
\bb
A_4 = \int_{\mathcal{M}_4} \varepsilon_{abcd} \,\omega_K^a \wedge \omega_K^b \wedge \omega_K^c \wedge \omega_K^d.
\ee
We left out $\varepsilon_{abcd} \, E^a \wedge E^b \wedge \omega_K^c \wedge \omega_K^d$, since it's a total derivative.

In addition to the above coset terms, there are Wess-Zumino terms \cite{Goon:2012dy}: they  appear
as pullbacks of certain five-forms on a four dimensional hyper-surface. Straightforward
calculations  show that only one of the possible
five WZ terms is independent of the coset terms already accounted above. This independent term reads as follows:
\bb\label{flat_WZ}
A_2\equiv \int_{\mathcal{M}_5} \varepsilon_{abcd} \omega_D \wedge E^a \wedge E^b \wedge \omega_K^c \wedge \omega_K^d\,.
\ee
The above expression for $A_2$  can be written more explicitly as follows:
\bb
A_2 = \int_{\mathcal{M}_4} \varepsilon_{abcd} \Big( \frac12 E^a \wedge E^b \wedge d \xi^c \wedge d \xi^d - \frac{2}{3} \xi^2 E^a \wedge E^b \wedge E^c \wedge d \xi^d  + \frac{1}{4} \xi^4 E^a \wedge E^b \wedge E^c \wedge E^d \Big).
\ee
As already noted in the previous section, due to the commutator $[K_a , P_b ] \propto - 2 \eta_{ab} D$,
not all the NG fields of the coset  are independent; the Inverse Higgs Constraint (IHC) \cite{Ivanov:1975zq}
can be invoked to eliminate  $\xi^a$, related to broken $K_a$,   in favor of a derivative of $\pi$,
related to broken $D$. Conventionally, this is done by imposing $\omega_D = 0$;
in our case, the IHC is a solution of the equations of
motion for $\xi^a$ \cite{McArthur:2010zm} that also satisfies $\omega_D = 0$:
\bb
\delta_\xi A_n = 0,~~n=0,1,2,3,4\,.
\ee
Hence, we use  IHC to express $\xi^\mu = - 1/2 ~e^{-\pi} ~\partial^\mu \pi$, and substitute this into the
actions $A_n, n=1,2,3,4$,  which then reduce to the conformal Galileons:
\bb\label{flat_S_0}
A_0 = \int d^4 x ~e^{4 \pi}\,,
\ee
\bb\label{flat_S_1}
A_1 \simeq \int d^4 x ~e^{ 2 \pi} \left( \partial \pi \right)^2,
\ee
\bb\label{flat_S_2}
A_2 \simeq \int d^4 x \left( \partial \pi \right)^2 \Big( L_1^{TD} + \frac12 \left( \partial \pi \right)^2 \Big),
\ee
\bb\label{flat_S_3}
A_3 \simeq \int d^4 x ~e^{ - 2 \pi} \left( \partial \pi \right)^2 \Big( L_2^{TD} - \frac{1}{2} \left( \partial \pi \right)^2
L_1^{TD} + \frac{1}{2}  \left( \partial \pi \right)^4 \Big),
\ee
\bb\label{flat_S_4}
A_4 \simeq \int d^4 x ~e^{ - 4 \pi} \left( \partial \pi \right)^2 \Big( L_3^{TD} - 3 \left( \partial \pi \right)^2 L_2^{TD} + 5 \left( \partial \pi \right)^4 L_1^{TD} - \frac{11}{4} \left( \partial \pi \right)^6 \Big)\,.
\ee
As already mentioned,   $L_n^{TD}$ denote the total derivative terms made of $\pi$ with  the
convention that $\left( \square \pi \right)^n$
enters with the unit coefficient, e.g., $L_1^{TD} = \square \pi$, $L_2^{TD} =( \square \pi)^2+...$, etc. Moreover,  each term in $A_n$ contains $2 n$ derivatives, but the number of fields is different in each of them. Each $A_n$ is invariant w.r.t. the conformal Galilean transformations (consisting of the linearly realized Poincar\'e transformations,  and non-linearly
realized special conformal transformations and dilatations of $\pi$). Equations of
motion for the Galileons  have at most two time derivatives acting per field.


\vskip 0.3in

\subsection{\large Gravitational Dressing of 4D Conformal Galileons}\label{CCZW_dyn_4d}

\vskip 0.1in

Any local flat-space CFT must be Weyl invariant after it is embedded covariantly
in  curved space-time \cite{Farnsworth:2017tbz}\footnote{One could of course break
explicitly Weyl symmetry by adding some breaking terms as long as they vanish in the flat space limit,
however, we will not be including  such terms and will  preserve Weyl invariance in the  classical action.}.
Thus, we expect  gravitationally dressed conformal Galileons to be
Weyl invariant. The Weyl transformations  of the relevant fields are \cite{Gabadadze:2018hos}:
\bb
e^a \rightarrow e^\sigma e^a,
\ee
\bb
\pi \rightarrow \pi - \sigma,
\ee
\bb
\xi^a \rightarrow \xi^a + \frac12 e^{-\pi} \partial^a \sigma\,.
\ee
Only three of the building blocks found in the previous section are invariant under these transformations:
the one forms $E^a$ and $\Omega_D$, and the two-form $\mathcal{R}^{ab}$.

It is instructive to recall how the construction works in a simpler, 3D case \cite{Gabadadze:2018hos}: the Weyl tensor vanishes identically in 3D and one can use the Schouten tensor,  $\mathcal{S}^a$, instead of the curvature two-form, $\mathcal{R}^{ab}$. Then,
out of  the  three one-forms $E^a, ~\Omega_D,$ and  $\mathcal{S}^a$, one can build four three-form actions for the 3D Galileons \cite{Gabadadze:2018hos}.

In 4D the construction  is more involved, the Weyl tensor is no longer zero and one has to use
the curvature two form, $\mathcal{R}^{ab}$,  as a building block. In addition,  one can also
define the Weyl invariant curvature one-form, $\mathcal{R}^a$,  and  a zero-form,
the Ricci scalar $\mathcal{R}$, by using the interior product:
\bb
\mathcal{R}^{a} = i_{E_a} \mathcal{R}^{ab} = e^{-\pi}  R^{b} + 4 \Omega_K^b + 2 \Omega_K E^b,
\5\5
\mathcal{R} = i_{E_b} i_{E_a} \mathcal{R}^{ab} = e^{-2 \pi}  R + 12 \Omega_K.
\ee
Here $E_a= E_a^\mu \partial_\mu $, with $E_a^\mu E^b_\mu = \delta^b_a$ and $\Omega_K \equiv i_{E_a} \Omega_K^a$.

Let us now build the action. At the level of zero derivatives we can write down only one term:
\bb
\int_{\mathcal{M}_4} \varepsilon_{a b c d}\, E^{a} \wedge  E^{b} \wedge E^{c} \wedge E^{d}.
\ee
This is the Weyl invariant version of the cosmological constant. In the unitary gauge, where $\pi=0$,
and in the metric formalism
\bb
\mathcal{A}_0 \equiv \int d^4x \sqrt{g}\,.
\label{A0}
\ee
At the level of two derivatives we can write down three invariants, but only one of them is independent:
\bb
 \int_{\mathcal{M}_4} \varepsilon_{a b c d} \mathcal{R}^{a b} \wedge E^{c} \wedge E^{d}.
\ee
This is a Weyl invariant extension of the Einstein-Hilbert (EH) term; Weyl symmetry can be fixed
by the unitary gauge,  $\pi=0$, reducing $\mathcal{A}_1$ to the EH term,
\bb
\mathcal{A}_1 \equiv \int d^4x \sqrt {g} R\,,
\label{A1}
\ee
(however one should keep in mind the Weyl anomaly, which we will discuss at the end of this section).

Furthermore, at the level of four derivatives there are three independent invariants:
\bb
\nonumber \int_{\mathcal{M}_4} \varepsilon_{a b c d} \mathcal{R}^{a b} \wedge \mathcal{R}^{c d},
\5\5
\int_{\mathcal{M}_4} \varepsilon_{a b c d} \mathcal{R}^{a} \wedge  \mathcal{R}^{b} \wedge E^{c} \wedge E^{d},
\5\5
\int_{\mathcal{M}_4} \varepsilon_{a b c d} E^{a} \wedge  E^{b} \wedge E^{c} \wedge E^{d} ~\mathcal{R}^2 .
\ee
Thus, at this level   one can  write a two parameter  action containing the above three terms. In general, the
conformal mode of this action will not be a Galileon, it would contain other high derivative terms.
To get the Galileon we'd need to tune the  two free parameters to one another in  such a way that the term
\bb
\nonumber 4 \int_{\mathcal{M}_4} \varepsilon_{a b c d} \Omega_K^{a} \wedge  E^{b} \wedge E^{c} \wedge E^{d}~ \Omega_K =
\int_{\mathcal{M}_4} \varepsilon_{a b c d} E^{a} \wedge  E^{b} \wedge E^{c} \wedge E^{d}~ \Omega_K^2,
\ee
has a vanishing coefficient. This constraint reduces the number of free parameters to one.
Furthermore,  using  IHC and  adopting  the unitary gauge, $\pi = 0$,
we can get the metric  form of the sought action:
\bb
\mathcal{A}_2 = \int d^4 x \sqrt{g} \Big[ \left( R_{\mu \nu \rho \sigma} R^{\mu \nu \rho \sigma}
- 4 \left[ R^2 \right] + R^2 \right) + \alpha W_{\mu \nu \rho \sigma} W^{\mu \nu \rho \sigma} \Big].
\label{GBW}
\ee
The term in the parenthesis is the GB term, it is a total derivative in 4D and
its integral is the Euler characteristics of the corresponding manifold.
The one parameter freedom in  (\ref {GBW}) enables us to  add
the  square of the Weyl tensor that has a trivial conformal structure,
and therefore its coefficient, $\alpha$,  is not fixed by our procedure.

At the level of six derivatives there are five independent terms that by naive counting of derivatives and
fields could potentially reduce  to conformal Galileons:
\begin{align}
\nonumber {} & \int_{\mathcal{M}_4} \varepsilon_{a b c d} \mathcal{R}^{a b} \wedge \mathcal{R}^{c d} ~\mathcal{R},
  \5
  \int_{\mathcal{M}_4} \varepsilon_{a b c d} \mathcal{R}^{a b} \wedge \mathcal{R}^{c} \wedge \mathcal{R}^d,
  \5
  \int_{\mathcal{M}_4} \varepsilon_{a b c d} \mathcal{R}^{a} \wedge \mathcal{R}^{b} \wedge \mathcal{R}^{c} \wedge E^d, \\
  \nonumber  {} & ~~~~~~~~~~ \int_{\mathcal{M}_4} \varepsilon_{a b c d} \mathcal{R}^{a} \wedge \mathcal{R}^{b} \wedge E^{c} \wedge E^d ~\mathcal{R},
  \5\5
  \int_{\mathcal{M}_4} \varepsilon_{a b c d} E^{a} \wedge E^{b} \wedge E^{c} \wedge E^d ~\mathcal{R}^3\,.
\end{align}
 The  action containing the above five terms has four independent parameters, besides its overall multiplier. Requiring that the action for the conformal mode  reduces to a conformal Galileon, we get two constraints on the 4 parameters, ensuring that  the following two terms have zero  coefficients:
\bb
\nonumber \int_{\mathcal{M}_4} \varepsilon_{a b c d} \Omega_K^{a} \wedge  \Omega_K^{b} \wedge E^{c} \wedge E^{d}~ \Omega_K,
\5\5
\int_{\mathcal{M}_4} \varepsilon_{a b c d} \Omega_K^{a} \wedge  E^{b} \wedge E^{c} \wedge E^{d}~ \Omega_K^2 .
\ee
Moreover, there are terms containing two powers of the curvature and two  covariant derivatives -- schematically  $D{\cal R}D{\cal R}$ and implying various contractions --  which would  in general give more
derivatives than the conformal Galileons have, hence we do not include them by fine tuning their coefficients
to zero. Furthermore,  there are other terms with two
covariant derivatives and two powers of curvature invariants arranged so that they would
vanish if restricted  to the conformal mode, e.g.,  $D{W}D{\cal R}$; such terms would
not modify the action for the conformal mode as a stand alone field, and will be included in
the full effective  theory.  They would give new  (super)-Planckian mass poles in the propagators
(see more discussions on this point in the next section).

Thus, we end up with a two parameter action at the level of six derivatives, which in the metric
form and in the unitary gauge can be written as follows:
\begin{align}\label{fixed_S_3}
\nonumber \mathcal{A}_3 = & \int d^4 x \sqrt{g} \Big[ - R_{\alpha \beta \mu \nu} R^{\alpha \beta \mu \nu } R
+ 12 {R_{\mu \nu}}^{\alpha \beta} R^\mu_\alpha R^\nu_\beta + 24 \left[ R^3 \right] - 24 R \left[ R^2 \right] +
4 R^3 + \\
{} & ~~~~~~~~~~~~~~ + \beta_1 W_{\alpha \beta \mu \nu} W^{\alpha \beta \mu \nu } R + \beta_2 {W_{\mu \nu}}^{\alpha \beta} R^\mu_\alpha R^\nu_\beta \Big].
\end{align}
As before, the terms proportional to the Weyl tensor are not uniquely determined by our procedure,
hence $\beta_1$ and $\beta_2$ are arbitrary real parameters. This gives a
 gravitationally dressed action for the $A_3$ conformal Galileon.

Last but not least, at the level of eight derivatives, there are six independent invariants which by
naive counting of derivatives and fields could potentially reduce  to conformal Galileons:
\begin{align}
\nonumber {} & \int_{\mathcal{M}_4} \varepsilon_{a b c d} \mathcal{R}^{a b} \wedge \mathcal{R}^{c d} ~\mathcal{R}^2,
  \5\5
  \int_{\mathcal{M}_4} \varepsilon_{a b c d} \mathcal{R}^{a b} \wedge \mathcal{R}^{c} \wedge \mathcal{R}^d ~\mathcal{R}, \\
  \nonumber  {} & \int_{\mathcal{M}_4} \varepsilon_{a b c d} \mathcal{R}^{a} \wedge \mathcal{R}^{b} \wedge \mathcal{R}^{c} \wedge \mathcal{R}^d,
  \5\5
  \int_{\mathcal{M}_4} \varepsilon_{a b c d} \mathcal{R}^{a} \wedge \mathcal{R}^{b} \wedge \mathcal{R}^{c} \wedge E^d ~\mathcal{R}, \\
  \nonumber {} & \int_{\mathcal{M}_4} \varepsilon_{a b c d} \mathcal{R}^{a} \wedge \mathcal{R}^{b} \wedge E^{c} \wedge E^d ~\mathcal{R}^2,
  \5\5
  \int_{\mathcal{M}_4} \varepsilon_{a b c d} E^{a} \wedge E^{b} \wedge E^{c} \wedge E^d ~\mathcal{R}^4\,,
\end{align}
requiring five independent parameters in the initial action.  In addition to these  terms there are ones
containing three powers of the curvature and two powers of a covariant derivative, or two powers of the curvature and
four powers of the covariant derivative. Some of these terms will be nonzero in the conformal limit
(i.e., in the limit when  only the conformal mode is kept)  and would have more derivatives than present in
the conformal Galileons -- we do not include such terms by tuning their coefficients to zero.
We will, however,  include in the full effective Lagrangian the terms that  vanish when restricted to
the conformal mode. Therefore, the full Lagrangian would still reduce to  the conformal Galileon
Lagrangian in the conformal limit (see next section).

Furthermore,  we impose three constraints on the parameters
to guarantee the absence of following terms
\begin{align}
\nonumber {} & \int_{\mathcal{M}_4} \varepsilon_{a b c d} \Omega_K^{a} \wedge  \Omega_K^{b} \wedge \Omega_K^{c} \wedge E^{d}~ \Omega_K,
\5\5
\int_{\mathcal{M}_4} \varepsilon_{a b c d} \Omega_K^{a} \wedge  \Omega_K^{b} \wedge E^{c} \wedge E^{d}~ \Omega_K^2 , \\
\nonumber {} & ~~~~~~~~~~~~~~~~~~~~~~~ \int_{\mathcal{M}_4} \varepsilon_{a b c d} \Omega_K^{a} \wedge  E^{b} \wedge E^{c} \wedge E^{d}~ \Omega_K^3 ,
\end{align}
and as a result end up with a two parameter action reproducing a conformal Galileon.
In the unitary gauge the action reads:
\begin{align}
\label{fixed_A_4}
\mathcal{A}_4 = & \int d^4 x \sqrt{g} \Big( \frac{1}{6} R_{\mu \nu \rho \sigma} R^{\mu \nu \rho \sigma} R^2 -4 {R_{\mu \nu}}^{\alpha \beta} R^\mu_\alpha R^\nu_\beta R
- 6 \left[ R^4 \right] + 3 \left[ R^2 \right]^2 +  \\
\nonumber & ~~~~~~~~~~~~~~~~~~~~  + \frac{8}{3} \left[ R^3 \right] R - \frac{1}{27} R^4
+ \gamma_1 W_{\mu \nu \rho \sigma} W^{\mu \nu \rho \sigma} R^2 + \gamma_2 {W_{\mu \nu}}^{\alpha \beta} R^\mu_\alpha R^\nu_\beta R \Big).
\end{align}
As before, there are two arbitrary real parameters, $\gamma_1$ and $\gamma_2$ not fixed
by our procedure.

To summarize so far the total coset action is:
\bb
{\cal A}_{\rm coset} = c_0{\cal A}_{0}+c_1 {\cal A}_{1} +c_2 {\cal A}_{2} + c_3 {\cal A}_{3}+c_4 {\cal A}_{4} ,
\label{Acoset}
\ee
with $c's$ being  real dimensionful coefficients,  $c_0$ setting the vacuum energy density and
$c_1$ defining the Planck scale square.  The initial Weyl invariance of the action, which
so far was gauge fixed  for simplicity, can easily be restored by a  substitution, $g \to g e^{2\pi}$.
While in the classical theory such field transformations are  harmless, this is not the case in the full
quantum theory due to the well known scale anomaly.

In that regard we note that, (\ref {GBW}) is not the only term in the second order in curvature; one can write a Wess-Zumino term:
\begin{align}
\nonumber
\mathcal{A}^{WZ}_2\equiv
\int_{\mathcal{M}_5} \varepsilon_{abcd} & \,  \Omega_D \wedge  \mathcal{R}^{ab}  \wedge \mathcal{R}^{cd} =
\int_{\mathcal{M}_4} \varepsilon_{abcd} \Big( 8 E^a \wedge E^b \wedge D \xi^c \wedge D \xi^d - \\
{} & - \frac{32}{3} \xi^2 ~ E^a \wedge E^b \wedge E^c \wedge D \xi^d + 4 \xi^4 ~ E^a \wedge E^b \wedge E^c \wedge E^d  + \\
\nonumber {} & + 8 E^a \wedge D \xi^b \wedge R^{cd} - 4 \xi^2 E^a \wedge E^b \wedge R^{cd}
+ \pi ~R^{ab} \wedge R^{cd} \Big{)}.
\end{align}
The equation of motion  for $\xi$ gives  IHC; substituting it  into $\mathcal{A}^{WZ}_2$
we recover  the effective action for the scale anomaly derived  in \cite{Riegert:1984kt} (see also
 \cite{Komargodski:2011vj},  where  the a-theorem has been proven and \cite{Bonifacio:2020vbk,Fernandes:2020nbq}, where it was derived from a different approach):
\begin{align}
\mathcal{A}^{WZ}_2 \simeq & \int d^4x \sqrt{g} \Bigg[ - \pi \left( R_{\mu \nu \rho \sigma} R^{\mu \nu \rho \sigma} - 4 \left[ R^2 \right] + R^2 \right) + \\
\nonumber {} & ~~~~~~~~~~~~~~~ + 4 \partial^\mu \pi \partial^\nu \pi \left( R_{\mu \nu} - \frac12 g_{\mu \nu} R \right) + 4 \left( \partial \pi \right)^2 \square \pi + 2 \left( \partial \pi \right)^4 \Bigg].
\end{align}
The latter action, unlike ${\cal A}_{\rm coset} (g e^{2\pi})$, is not Weyl invariant. Its Weyl transformation
gives a functional  the variation of which, taken at $\pi=0$, gives the  trace anomaly.\footnote{
 We note that the analogous considerations in  2D would have given  a  3D  WZ term:
\begin{align}
\int_{\mathcal{M}_3} \varepsilon_{ab} \,  \Omega_D \wedge & \mathcal{R}^{ab}
\simeq \int d^2x \sqrt{g} \Bigg[ - \pi R -  \left( \partial \pi \right)^2 \Bigg],
\nonumber
\end{align}
which reduces to  the 2D Polyakov action. }

Thus, the total action of the theory is ${\cal A}_{\rm total}$  in which the $\pi$ field  should be
restored via the substitution, $g \to g e^{2\pi}$,  plus the anomalous action $\mathcal{A}^{WZ}_2$
 \bb
{\cal A}_{\rm total} = {\cal A}_{\rm coset}(ge^{2\pi}) + c_{WZ}\mathcal{A}^{WZ}_2(g, \pi)\,,
\label{Atotal}
\ee
with ${\cal A}_{\rm coset}$  defined in (\ref {Acoset}).  The value of the constant $c_{WZ}$  is in general
determined by the number of degrees of freedom  coupled to gravity, and since we have not introduced any
matter degrees of freedom in our case it's only $\pi$ that contributes to $c_{WZ}$ .
As noted above, the $\pi$ field can no longer be gauged away, because of the anomalous term. If one were to introduce additional fields in the above action,  they would couple to gravity via the Weyl invariant couplings, i.e., they'd couple to $ge^{2\pi}$.

The flat space Conformal  Gallileons can be recovered from ${\cal A}_{\rm total}$  via the substitution
$g_{\mu\nu} = \eta_{\mu\nu}$:  the term ${\cal A}_{\rm coset}(\eta \,e^{2\pi})$ produces a weighted sum of the
$A_0,A_1,A_3,A_4$  Conformal Galileons of the previous subsection, while $\mathcal{A}^{WZ}_2(\eta, \pi)$
yields  the $A_2$ Conformal Galileon, which was obtained as a WZ term in the flat space case \cite{Goon:2012dy}.

\vskip 0.3in

\section{Effective Field Theory and Cosmology}\label{sec_stability}

\vskip 0.1in


 The key property of the action (\ref {Atotal}) is that the equation of motion
 -- when  restricted to the conformal mode --  has no more than two derivatives
 acting on each field,  and hence, no new degrees of freedom
 emerge in the conformal sector.

 This is  important for cosmology, where the background evolution of a homogeneous and
 isotropic universe is described by the conformal mode, the scale  factor.

 However,  the existence of the higher powers of the curvatures in the action suggests that the
 tensor mode propagator will in general have additional poles on curved backgrounds.
 Such poles, in a fundamentally Lorentz invariant theory,  describe  ghosts  that invalidate the theory,
 unless the mass of these modes are above the
 cutoff of the effective field theory (EFT) that (\ref {Atotal}) is part of.

Indeed, the potentially problematic  modes  in (\ref {Atotal})  have in general
a super-Planckian masses  as long as  the background curvatures are sub-Planckian
(this feature is not necessarily specific to the choice of the coefficients between the various
terms in (\ref {Atotal})). To see this,
consider  perturbations $h_{\mu \nu} = g_{\mu \nu} - g^b_{\mu \nu}$,
where $g^b$  stands for the background metric. For simplicity, we will focus only on
the $R^3$ terms,  ignore all the tensorial indices,  and  assume  that the background
curvature, $R_b$, is constant. Then, in the leading order the Lagrangian for small perturbations
within any locally flat small neighborhood of a space-time point  would  take the following
schematic form
\bb
\mathcal{L} =  h \partial^2 h + \frac{ c R_b }{ M_p^{4}}
\Big( h \partial^4 h +  R_b h \partial^2 h \Big)\,.
\ee
The inverse of the propagator  for $h$ would  read, $
{p^2  ( 1 + ({ c R_b^{2} }/{  M_p^{4}}) -  ({  c R_b  p^2}/{  M_p^{4}}))},$
which guarantees a massless  and massive poles in the propagator.
In a fundamentally Lorentz invariant theory that we are dealing with the
massive pole is necessarily a ghost if the massless one is not. For reasonable values
of the parameter $ c \sim {\cal O}(1)$,  and  for a sub-Planckian background, $R_b<<M_p^2$,
the ghost mass is super-Planckian,  $M_{ghost}^2 \simeq M_p^2  ({ M_p^{2}}/{ c R_b }).$
and this is outside the effective field theory regime.

 On the other hand,  if the action (\ref {Atotal}) is just part of  an EFT
 with an infinite number of other terms of growing dimensionality,
 then the additional terms may spoil the key property of the conformal mode in (\ref {Atotal}) --
 the facts that this mode is a conformal Galileon. There is a way to deal with
 this issue as discussed below.

 Since the higher order conformal Galileons do not exist,  the conformal structure would be maintained
 only if all the additional terms in EFT vanish  for the conformal mode.
  One way to realize this is to have all the higher order terms be proportional to at least one
  power of the  Weyl tensor.\footnote{The trace of the variation
  of the Weyl tensor is proportional to the Weyl tensor, so one
  power  on $W$ is  already acceptable for our goals.} In this case,  the total effective field theory action  would take the form:
 \bb
{\cal A}_{\rm EFT} = {\cal A}_{\rm total} + \int d^4x \sqrt{\bar g} \sum_{k=1}^{\infty}\, \sum_{l=0}^{\infty}\,
\sum_{m=1}^{\infty} b_{klm}{W}^k {\bar D}^l {\bar R}^m\,,
\label{A_EFT}
\ee
where $b_{klm}$ are some coefficients, ${\bar g} =g e^{2\pi}$,  ${W}^k$ denotes powers of the
Weyl tensor,  and ${\bar R}$  and $\bar D$ denote respectively
the Riemann tensor and covariant derivative for $\bar g$,  with all possible
contractions done by the inverse of the metric $\bar g$.  Terms such as $W{\bar D}^2 {\bar R}$
would introduce new poles on the flat space, but their masses are (super)-Planckian
for reasonable values of the  parameters in front of such terms.

 In each order of EFT in (\ref {A_EFT}) there could in general be other terms with more derivatives, such as,
 ${\bar R}{\bar D}^2{\bar R}$, or ${\bar R}^2{\bar D}^2{\bar R}$, and so on;
 the coefficients of  such terms have been set to zero to guarantee the
 properties  of the conformal sector that we desired. We are not aware of a universal
 principle that would guarantee  such cancellations in the full quantum theory.
 Such a principle could have emerged  due to a theory that would complete
 the present one at  and above the Planck scale. Until that theory is known, our
 procedure should be regarded as a order-by-order fine tuning of the coefficients of
 counter-terms to render the renormalized EFT action free of the higher
 derivative terms, when it reduced to the theory of the conformal mode only.

 The conformal mode is not a physical propagating mode in the EH action.
 Moreover, its kinetic term has a ghost sign. Nevertheless, it is the mode that describes
 evolution of the Friedmann-Lemaitre-Robertson-Walker (FLRW)  universe. Making sure
 that no extra derivatives emerge in this sector, i.e., that no Ostrogradski instabilities appear for
 the conformal mode, is the first stepping stone toward a potentially viable cosmology in any theory
 with additional terms in the action.

 Equipped with the above knowledge we can briefly consider  cosmology in a mini-superspace approximation
simply to see what novel features  might  be introduced by the higher curvature terms, and how those novelties
play out in the context of EFT.  In what follows in this section we set $\pi=0$ as an {\it ansatz},
and assume the FLRW  metric
\bb
g^{FLRW}_{\mu \nu} dx^\mu dx^\nu =  \frac{1}{n^2 (t)} dt^2 - \frac{a^2 (t)}{1- k r^2} dr^2  - a^2 (t) r^2 d\theta^2 -
a^2 (t) r^2 \sin^2 \theta d\phi^2 \,.
\ee
These substitutions nullify all the terms in the EFT action (\ref {A_EFT}) that are proportional to the Weyl tensor.
Furthermore, all the curvature invariants  are  straightforwardly  expressible in terms of the scale factor, $a$, and
the inverse lapse, $n$.   What remains of the full EFT action (\ref {A_EFT}) is the following expression
\bb
- \frac{1}{12} \int d^4 x \sqrt{g^{FLRW}}  R(g^{FLRW})  - \frac{1}{24 \beta} \mathcal{A}_3(g^{FLRW}) - \frac{1}{48 \gamma} \mathcal{A}_4(g^{FLRW})\,,
\label{Cosmo}
\ee
where we also set the cosmological constant to zero, and renamed the arbitrary coefficients
in front of the cubic and quartic order terms as $\beta$ and $\gamma$.
The respective mini-superspace Lagrangian reads
(up to a total derivative)
\begin{align}
\nonumber  & \frac{V}{2} \Bigg[ \frac{a}{n} \left( k - n^2 \dot{a}^2 \right)  -
\frac{1}{a^3 n \beta} \left( k + n^2 \dot{a}^2 \right)^3
+ \frac{6  }{a^3 \beta} \left( k^2 \dot{a}^2 n + \frac{2}{3} k \dot{a}^4 n^3 + \frac{1}{5} \dot{a}^6 n^5 \right) +
\\
{} & ~~~~~ + \frac{1}{n a^5 \gamma} (k+n^2 \dot{a}^2)^4
- \frac{8 }{a^5 \gamma} \left( k^3 n \dot{a}^2 + k^2 n^3 \dot{a}^4 + \frac{3}{5} k
n^5 \dot{a}^6 + \frac{1}{7} n^7 \dot{a}^8 \right) \Bigg]\,.
\end{align}
Here $V$ stands for the volume of the space: $V=2 \pi^2$ for $k=1$,
$V=\infty$ for $k=0$ and $k=-1$. By construction, there are no second and higher time derivatives
of the scale factor appearing in the Lagrangian above. Furthermore, let us introduce the notations:
\bb
H \equiv \frac{n\dot{a}}{a}\,,
\5\5\5
y \equiv H^2 + \frac{k}{a^2 (t)}\,.
\ee
Then, the modified Friedmann  equation for an empty space-time reads:
\bb
y  - \frac{1}{\beta} y^3 + \frac{1}{\gamma} y^4  = 0\,.
\label{MFriedmann}
\ee
The cubic and quartic terms in $y$ are suppressed by  the respective  powers of the
cutoff scale (Planck scale). Thus,  these terms  would modify  conventional
solutions of the ordinary  Friedmann equation  by small corrections,
as long as the physical scales involved in those solution are significantly lower
than the Planck mass scale.

Note however that there are new solutions to the modified Friedmann equation (\ref {MFriedmann})
which do not exist for the conventional equation. Such solutions could  be looked for
by  finding the  zeros of the spacial quartic polynomial in (\ref {MFriedmann}).
To see explicitly some of these  solutions
let us drop the quartic term in $y$  by taking the limit, $ \gamma \rightarrow \infty$.
Then, putting $k=1$ and $n=1$ one  finds a Starobinski-like solution for a closed universe,
which in this case describes a contracting and then  re-expanding universe:
\bb
a (t) = {1\over {\beta}^{1/4}}\, \cosh \left( \beta^{1/4} t \right)\,.
\label{Starobinski}
\ee
There is a similar solution for  a spatially flat,  as well as open universes, all three
representing the de Sitter space-time. Furthermore, there is also a static solution
corresponding to the negative root of the quadratic equation, $y^2= {\beta}$,
with an open spatial section, $k=-1$, and a constant
scale factor $a=1/\beta^{1/4}$. Weather any of  these solutions can be stable with respect to small
perturbations is an interesting open question.  We  only point out  that all these solutions invoke
curvatures  at the cutoff  of  EFT and are likely to be strongly modified if one were to include other
derivative terms in the same order that do not reduce to conformal Galileons.
Therefore, the above solutions, and their extensions, require certain fine tunings of the parameters,
as  it was done in (\ref {A_EFT}).


\vskip 0.3in

\section{\Large Lower Dimensional Descendants of Lovelocks}\label{sec_main}

\vskip 0.1in


In this section we will  re-derive the coset action from the higher dimensional Lovelock
terms, by using certain identities \cite{DeWitt:1965jb,1970PCPS...68..345L}. This method allows
us to see familiar results from a different perspective.


\subsection{\large 3D Example}


In \cite{Gabadadze:2018hos} we showed that $3D$ conformal Galileon could be viewed as the
St\"uckelberg field restoring the local Weyl symmetry in NMG \cite{Bergshoeff:2009hq},
and its extension \cite {Sinha:2010ai}.  Here, we will obtain the same
3D action by dimensionally reducing higher dimensional Lovelock terms.
The next subsection will deal with the  4D case.

Let us begin with  the following four Lovelock expressions in space-time of dimensionality
$n\ge 3$, which will be relevant for our construction in $3D$:
\bb
\mathcal{A}^{nD}_0 = - \int_{\mathcal{M}_n} \varepsilon_{a_1 \cdots a_n} E^{a_1} \wedge \cdots \wedge E^{a_n}\,,
\ee
\bb
\mathcal{A}^{nD}_1 = - \int_{\mathcal{M}_n} \varepsilon_{a_1 \cdots a_n} E^{a_1} \wedge \cdots \wedge E^{a_{n-2}} \wedge \mathcal{R}^{a_{n-1} a_n}\,,
\ee
\bb\label{unreg_cov_S_2_nd}
\mathcal{A}^{nD}_2 = - \int_{\mathcal{M}_n} \varepsilon_{a_1 \cdots a_n} E^{a_1} \wedge \cdots \wedge E^{a_{n-4}} \wedge \mathcal{R}^{a_{n-3} a_{n-2}} \wedge \mathcal{R}^{a_{n-1} a_n}\,,
\ee
\bb\label{unreg_cov_S_3_nd}
\mathcal{A}^{nD}_3 = - \int_{\mathcal{M}_n} \varepsilon_{a_1 \cdots a_n} E^{a_1} \wedge \cdots \wedge E^{a_{n-6}} \wedge \mathcal{R}^{a_{n-5} a_{n-4}} \wedge \mathcal{R}^{a_{n-3} a_{n-2}} \wedge \mathcal{R}^{a_{n-1} a_n}\,.
\ee
These  also happen to be the terms of the Conformal/Poincar\'e coset, hence they
realize nonlinearly the  special conformal and dilatation symmetries.  Moreover,
all of these terms have the desired  conformal structure, i.e., the conformal mode is a conformal
Galileon in $n$ dimensional space-time. The IHC is a solution to the corresponding
equations of motion, so we will use $\xi^\mu = -1/2 e^{-\pi} \partial^\mu \pi$.
(It is convenient to work in the unitary gauge $(\pi=0)$, and if needed,
one can recover the $\pi$ interactions by making a field redefinition  $g_{\mu \nu} \rightarrow e^{2\pi} g_{\mu \nu}$ in
the unitary gauge classical action.)

Let us rewrite the above  expressions in the metric form:
\bb\label{unreg_cov_S_0_3d}
\mathcal{A}^{nD}_0 = n!\,,
\ee
\bb\label{unreg_cov_S_1_3d}
\mathcal{A}^{nD}_1 = \int d^n x \sqrt{g} \Big[ \frac{1}{2} (n-2)! \delta^{\mu_1 \mu_2}_{\nu_1 \nu_2} {R^{\nu_1 \nu_2}}_{\mu_1 \mu_2} \Big]\,,
\ee
\bb\label{unreg_cov_S_2_3d}
\mathcal{A}^{nD}_2 =
\int d^n x \sqrt{g} \Big[ \frac{1}{2^2} (n-4)! \delta^{\mu_1 \cdots \mu_4}_{\nu_1 \cdots \nu_4} {R^{\nu_1 \nu_2}}_{\mu_1 \mu_2} {R^{\nu_3 \nu_4}}_{\mu_3 \mu_4} \Big]\,,
\ee
\bb\label{unreg_cov_S_3_3d}
\mathcal{A}^{nD}_3 = \int d^n x \sqrt{g} \Big[ \frac{1}{2^3} (n-6)! \delta^{\mu_1 \cdots \mu_6}_{\nu_1 \cdots \nu_6}
{R^{\nu_1 \nu_2}}_{\mu_1 \mu_2} {R^{\nu_3 \nu_4}}_{\mu_3 \mu_4} {R^{\nu_5 \nu_6}}_{\mu_5 \mu_6} \Big]\,.
\ee
Note that $\mathcal{A}^{nD}_0$ and $\mathcal{A}^{nD}_1$ are both regular  for $n=3$, the problems arise with $\mathcal{A}^{nD}_2$ and $\mathcal{A}^{nD}_3$, both of  which contain products $\infty \times 0$. The infinity comes
from the factorial and zero from the generalized Kronecker symbol  (the latter is zero in $n=3$ because it contains more than three antisymmetrised indices). One needs to regularize the above expressions to make sense of them.
Regularization of  $\mathcal{A}^{nD}_2$ is relatively easy, one can assume that for $n<4$ the Weyl tensor
${W^{\mu \nu}}_{\alpha \beta}$ vanishes identically;  obviously,  the Weyl tensor vanishes  for $n=3$,
but the assumption is that  it also vanishes for $3\leq n<4$ as we analytically continue the parameter $n$.
This enables us to express the Riemann tensor   for $3\leq n<4$ as follows:
\bb\label{weyl_zero}
{R^{\mu \nu}}_{\alpha \beta} = \frac{1}{n-2} \Big( R^\mu_\alpha \delta^\nu_\beta - R^\mu_\beta \delta^\nu_\alpha + \delta^\mu_\alpha R^\nu_\beta - \delta^\mu_\beta R^\nu_\alpha \Big) - \frac{1}{\left( n-2 \right) \left( n-1 \right)} R
\Big( \delta^\mu_\alpha \delta^\nu_\beta - \delta^\mu_\beta \delta^\nu_\alpha \Big)\,.
\ee
We can now substitute this relation  into  the Lovelock  action  (\ref{unreg_cov_S_2_3d})  to express
the latter as follows:
\bb
\mathcal{A}^{nD}_2 =
\int d^n x \sqrt{g} \Big[ -4  \frac{(n-3)!}{n-2} \Big( \left[ R^2 \right] - \frac{n}{4 \left( n-1 \right)} R^2 \Big) \Big]\,.
\ee
This expression is regular for $n=3$, and upon this substitution
gives the  action of NMG \cite{Bergshoeff:2009hq}.

The above procedure -- referred as the method of higher dimensional reduction (HDR) --
defines formal analytic continuation of Lovelocks
from $n>3$, down to $n=3$.  We will use now  HDR  for other terms.
\footnote {An equivalent method of regularization in this case is to subtract the term
$\delta^{\mu_1 \cdots \mu_4}_{\nu_1 \cdots \nu_4} {W^{\nu_1 \nu_2}}_{\mu_1 \mu_2} {W^{\nu_3 \nu_4}}_{\mu_3 \mu_4}$
from the original $n$ dimensional action with an appropriate coefficient to eliminate the maximal tensorial structure (in this case $R_{\alpha \beta \mu \nu} R^{\alpha \beta \mu \nu}$). The two methods are equivalent.
It turns out that in more complicated cases the latter method is more convenient;  also the
subtraction does explain the emergence of an additional massive spin-2 mode in the lower dimensional
actions, while this mode does not exist in the starting Lovelock action.}

To regularize ${\cal A}_3^{nD}$ we have to take more steps down the ladder of dimensionalities.
First we regularize (\ref{unreg_cov_S_3_3d}) in $5D$, then in $4D$, and only after  in $3D$;
hence we have a cascade of regularizations
\bb
\nonumber 6D \rightarrow 5D \rightarrow 4D \rightarrow 3D\,.
\ee
The first step in this cascade consists of adopting  the following identity for $n<6$:
\bb
\delta^{\mu_1 \cdots \mu_6}_{\nu_1 \cdots \nu_6} {W^{\nu_1 \nu_2}}_{\mu_1 \mu_2} {W^{\nu_3 \nu_4}}_{\mu_3 \mu_4} {W^{\nu_5 \nu_6}}_{\mu_5 \mu_6} = 0\,.
\ee
As soon as this is used,  the action  (\ref{unreg_cov_S_3_3d}) for $\mathcal{A}^{nD}_3$ takes the form
\begin{align} \nonumber
\int d^n x \sqrt{g} \frac{(n-5)!}{n-2} \Big[ \frac{3 (n+2)}{n-1} {R_{\alpha \beta}}^{\mu \nu} {R_{\mu \nu}}^{\alpha \beta} R
+ 24 {R_{\mu \nu}}^{\alpha \beta} {R_{\beta \rho}}^{\mu \nu} R^\rho_\alpha + \frac{24 n }{n-2} {R_{\mu \nu}}^{\alpha \beta} R^\mu_\alpha R^\nu_\beta + \\ \nonumber \frac{16n (n-1)}{(n-2)^2} \left[ R^3 \right] - \frac{12 (n^3 - 2 n^2 + 6 n - 8)}{(n-1) (n-2)^2} \left[ R^2 \right] R +  \frac{n^4 - 3 n^3 +10 n^2 +4 n -24}{(n-1)^2 (n-2)^2} R^3 \Big]\,,
\end{align}
which is now regular in $5D$. The next step is to descend to $4D$.  The expression in the square brackets
is zero in $n=4$, while the overall multiplier is diverging, hence further regularization is needed.
The latter can be  achieved by means of the following  identity:
\bb\label{weyl_square_tensor}
R^\mu_\nu\, \delta_{\mu \alpha \beta \gamma \delta}^{\nu \lambda \rho \sigma \omega}  {W_{\lambda \rho}}^{\alpha \beta} {W_{ \sigma \omega }}^{\gamma \delta} = R^\mu_\nu\, \cdot 0 =0\,.
\ee
Using the above, the regularized $\mathcal{A}^{nD}_3$  for $n< 5$ can be written  as follows:
\begin{align}\label{S_3_reg_4D}
\mathcal{A}^{nD}_3 = & \int d^n x \sqrt{g} \frac{(n-4)!}{n-2} \Big[ - \frac{3}{n-1} R_{\alpha \beta \mu \nu} R^{\alpha \beta \mu \nu } R
+ \frac{24 }{n-2} {R_{\mu \nu}}^{\alpha \beta} R^\mu_\alpha R^\nu_\beta + \\
\nonumber {} & + \frac{16n }{(n-2)^2} \left[ R^3 \right] -
\frac{12 n^2}{(n-1) (n-2)^2} \left[ R^2 \right] R +
\frac{n (n^2 + n + 2)}{(n-1)^2 (n-2)^2} R^3 \Big]\,.
\end{align}
It is straightforward to check that the  action is regular  for $n=4$. For the last
step of the cascade we use (\ref{weyl_zero})  to get a perfectly regular
expression for $n<4$:
\bb
\mathcal{A}^{nD}_3 = \int d^n x \sqrt{g} \frac{(n-3)!}{(n-2)^3} \Big[  16 \left[ R^3 \right] -
\frac{12 n}{n-1} \left[ R^2 \right] R +
\frac{n^2 + 4n - 4}{(n-1)^2} R^3 \Big]\,.
\ee
Substituting  $n=3$ in the above expression we  get the action that extends
NMG to the cubic order \cite{Sinha:2010ai}. The fact  that  there is a connection
between  the Lovelock terms and NMG was already shown  in Ref.
\cite{Sinha:2010ai}  on the basis of equivalence between their dual CFTs (see also \cite{Oliva:2010eb,Alkac:2020zhg}).
Our method is complementary.

\vskip 0.3in

\subsection{\large 4D Theory}

\vskip 0.1in

By now one should perhaps expect that the gravitationally  dressed conformal
Galileons are certain descendants of  higher dimensional Lovelock terms.
This point of view will be reinforced below by obtaining
the coset action of Section 3  from the Lovelock terms,
using the method of HDR.

Let us first briefly summarize the rules of HDR:
\begin{itemize}
	\item To descend from $m$ dimensions down to $(m-1)$ dimensions we use
	identities  involving  the symbol $\delta^{\mu_1 \cdots \mu_m}_{\nu_1 \cdots \nu_m}$
	contracted with  curvature  tensors.
	\item Each identity must involve at least one Weyl tensor; otherwise the identity should not
	use  the Riemann tensor.
	\item At every step there are a finite number of the identities to be used,  and usually one  needs all of them.
\end{itemize}
After using  the identities one gets $(m-1) \times {L}^k$, where  ${L}^k$ is the analogue of the
$k$th order Lovelock in $(m-1)<2k$ dimensions (i.e., it has the same conformal structure as the Lovelock).
If the rules described above are not sufficient to extract the factor of $(m-1)$, then the corresponding Lovelock does
not have an analogue in $(m-1)$ dimensions (i.e., there are no terms  with the conformal structure of the  Lovelock). Note that this definition of an ``analogue" leaves a room for degeneracy (as we will see below), but the
terms we get through the above described procedure are guaranteed to be the most general ones.

Let us now apply HDR in 4D. Among the terms (\ref{unreg_cov_S_0_3d})-(\ref{unreg_cov_S_3_3d})
 only $\mathcal{A}^{nD}_3$ needs regularization in 4D, however, this  was already done in
 (\ref{S_3_reg_4D}) {\it en~route} to the $3D$ expression. Hence, substituting  $n=4$  into
 (\ref{S_3_reg_4D})  we get:
\bb\label{S_3_4D}
\mathcal{A}^{4D}_3 = \int d^4 x \sqrt{g}  \frac{1}{2} \Big( - R_{\alpha \beta \mu \nu} R^{\alpha \beta \mu \nu } R
+ 12 {R_{\mu \nu}}^{\alpha \beta} R^\mu_\alpha R^\nu_\beta + 16 \left[ R^3 \right] - 16 \left[ R^2 \right] R + \frac{22}{9} R^3 \Big).
\nonumber
\ee
This action coincides with  $\mathcal{A}_3 $ that we've derived in  (\ref{fixed_S_3}) by using  the coset
construction\footnote{It corresponds to the case of  $\beta_1=\beta_2 =0$,  but the latter terms can always be
added since they do not affect the conformal structure,  and are part of  the degeneracy we
mentioned in the previous paragraph.}.

Let us now see how $\mathcal{A}_4 $ of (\ref {fixed_A_4}) comes along in this formalism. For this we look
at an eight derivative Lovelock term in  $n$ dimensions:
\bb\label{unreg_cov_S_4_nd}
\mathcal{A}^{nD}_4 = - \int_{\mathcal{M}_n} \varepsilon_{a_1 \cdots a_n} E^{a_1} \wedge \cdots \wedge E^{a_{n-8}} \wedge \mathcal{R}^{a_{n-7} a_{n-6}} \wedge \mathcal{R}^{a_{n-5} a_{n-4}} \wedge \mathcal{R}^{a_{n-3} a_{n-2}} \wedge \mathcal{R}^{a_{n-1} a_n}.
\ee
In the unitary gauge  and in the metric formulation the above  expression  becomes:
\bb
\mathcal{A}^{nD}_4 = \int d^n x \sqrt{g} \frac{1}{2^4} (n-8)! \delta^{\mu_1 \cdots \mu_8}_{\nu_1 \cdots \nu_8}
{R^{\nu_1 \nu_2}}_{\mu_1 \mu_2} {R^{\nu_3 \nu_4}}_{\mu_3 \mu_4} {R^{\nu_5 \nu_6}}_{\mu_5 \mu_6} {R^{\nu_7 \nu_8}}_{\mu_7 \mu_8}\,.
\ee
Regularization of this expression is a tedious task. We will not fully describe the process but rather give the necessary identities for each step of HDR:
\begin{align}
\nonumber 8D \rightarrow 7D & \5\5\5\5\5 \delta^{\mu_1 \cdots \mu_8}_{\nu_1 \cdots \nu_8} {W^{\nu_1 \nu_2}}_{\mu_1 \mu_2} {W^{\nu_3 \nu_4}}_{\mu_3 \mu_4} {W^{\nu_5 \nu_6}}_{\mu_5 \mu_6} {W^{\nu_7 \nu_8}}_{\mu_7 \mu_8} = 0,\\
\nonumber 7D \rightarrow 6D & \5\5\5\5\5 \delta^{\mu_1 \cdots \mu_7}_{\nu_1 \cdots \nu_7} {W^{\nu_1 \nu_2}}_{\mu_1 \mu_2} {W^{\nu_3 \nu_4}}_{\mu_3 \mu_4} {W^{\nu_5 \nu_6}}_{\mu_5 \mu_6} R^{\nu_7}_{\mu_7} = 0, \\
\nonumber 6D \rightarrow 5D & \5\5\5\5\5  \delta^{\mu_1 \cdots \mu_6}_{\nu_1 \cdots \nu_6} {W^{\nu_1 \nu_2}}_{\mu_1 \mu_2} {W^{\nu_3 \nu_4}}_{\mu_3 \mu_4} {W^{\nu_5 \nu_6}}_{\mu_5 \mu_6} R =0, \\
\nonumber {} & \5\5\5\5\5 \delta^{\mu_1 \cdots \mu_6}_{\nu_1 \cdots \nu_6} {W^{\nu_1 \nu_2}}_{\mu_1 \mu_2} {W^{\nu_3 \nu_4}}_{\mu_3 \mu_4} {R^{\nu_5}}_{\mu_5} {R^{\nu_6}}_{\mu_6} = 0, \\
\nonumber 5D \rightarrow 4D & \5\5\5\5\5 \delta^{\mu_1 \cdots \mu_5}_{\nu_1 \cdots \nu_5} {W^{\nu_1 \nu_2}}_{\mu_1 \mu_2} {W^{\nu_3 \nu_4}}_{\mu_3 \mu_4} {R^{\nu_5}}_{\mu_5} R = 0, \\
\nonumber & \5\5\5\5\5 \delta^{\mu_1 \cdots \mu_5}_{\nu_1 \cdots \nu_5} {W^{\nu_1 \nu_2}}_{\mu_1 \mu_2} {R^{\nu_3}}_{\mu_3} {R^{\nu_4}}_{\mu_4} {R^{\nu_5}}_{\mu_5} = 0.
\end{align}
The result of this lengthy procedure is the regularized action  valid for $4\leq n<5$:
\begin{align}
\nonumber \mathcal{A}^{nD}_4 = & \int d^n x \sqrt{g} \frac{(n-4)!}{(n-1)^3 (n-2)^4} \Big[ 6 (n-2)^2 (n-1) R_{\mu \nu \rho \sigma} R^{\mu \nu \rho \sigma} R^2 - \\
\nonumber & ~~ -96 (n-2) (n-1)^2 {R_{\mu \nu}}^{\alpha \beta} R^\mu_\alpha R^\nu_\beta  R
-96 (n-1)^3 \left[ R^4 \right] + 48 (n-1)^3 \left[ R^2 \right]^2 + \\
{} & ~~ + 64 (n-2) (n-1)^2 \left[ R^3 \right] R -24 n (n-4) (n-1) \left[ R^2 \right] R^2 + \\
\nonumber & ~~ + \left( n^3+11 n^2-78 n+56 \right) R^4 \Big]\,.
\end{align}
Substituting  $n=4$ into the above expression we get the $4D$ action in the unitary gauge:
\begin{align}\label{S_4_4D}
\mathcal{A}^{4D}_4 = & \int d^4 x \sqrt{g} \Big( \frac{1}{6} R_{\mu \nu \rho \sigma} R^{\mu \nu \rho \sigma} R^2 -4 {R_{\mu \nu}}^{\alpha \beta} R^\mu_\alpha R^\nu_\beta R  -
\\ \nonumber & ~~~~~~~~~~~~~~~~~~~~
- 6 \left[ R^4 \right] + 3 \left[ R^2 \right]^2  + \frac{8}{3} \left[ R^3 \right] R - \frac{1}{27} R^4  \Big)\,.
\end{align}
This coincides with $\mathcal{A}_4 $ in (\ref {fixed_A_4}),   with $\gamma_1=\gamma_2 =0$. This completes
the  derivation of   all the 4D coset   terms from the higher dimensional Lovelock terms.



\section*{Acknowledgements}

We'd  like to thank Massimo Porrati, Samson Shatashvili, Jim Simons,  and Alex Vilenkin for useful discussions.
The work  on GG is supported by  the NSF grant PHY-1915219.   GT is supported by the Simons Foundations
``Origins of the Universe" program.



\end{document}